\documentclass[conference]{IEEEtran}
\IEEEoverridecommandlockouts
\usepackage{cite}
\usepackage{amsmath,amssymb,amsfonts}
\usepackage{algorithmic}
\usepackage{graphicx}
\usepackage{textcomp}
\usepackage{xcolor}
\usepackage{multirow}
\usepackage{threeparttable}
\usepackage{url}
\usepackage{enumitem}
\usepackage{hyperref}
\usepackage{booktabs}
\def\BibTeX{{\rm B\kern-.05em{\sc i\kern-.025em b}\kern-.08em
    T\kern-.1667em\lower.7ex\hbox{E}\kern-.125emX}}
\begin{document}

\title{UniSep: Universal Target Audio Separation with Language Models at Scale}

\author{
\IEEEauthorblockN{
Yuanyuan Wang$^{1}$, Hangting Chen$^{2}$, Dongchao Yang$^1$, Weiqin Li$^{3}$, Dan Luo$^{3}$, \\ Guangzhi Li$^{2}$, Shan Yang$^{2}$, Zhiyong Wu$^{1,3}$, Helen Meng$^{1}$, Xixin Wu$^{1,\dagger}$
}

\thanks{Demo can be found in \href{https://uniseparation.github.io/UniSep/}{https://uniseparation.github.io/UniSep/}. $^\dagger$Corresponding author. This work is partially supported by the General Research Fund from the Research Grants Council of Hong Kong SAR Government (Project No. 14202623) and the Center for Perceptual and Interactive Intelligence (CPII) Ltd under the Innovation and Technology Commission’s InnoHK Scheme.}

\IEEEauthorblockA{$^1$ The Chinese University of Hong Kong, Hong Kong SAR, China} 

\IEEEauthorblockA{$^2$ Tencent AI Lab, Audio and Speech Signal Processing Oteam, China}

\IEEEauthorblockA{$^3$ Shenzhen International Graduate School, Tsinghua University, Shenzhen, China} 

}

\maketitle

\begin{abstract}
We propose Universal target audio Separation (UniSep), addressing the separation task on arbitrary mixtures of different types of audio. 
Distinguished from previous studies, UniSep is performed on unlimited source domains and unlimited source numbers.
We formulate the separation task as a sequence-to-sequence problem, and a large language model (LLM) is used to model the audio sequence in the discrete latent space, leveraging the power of LLM in handling complex mixture audios with large-scale data. 
Moreover, a novel pre-training strategy is proposed to utilize audio-only data, which reduces the efforts of large-scale data simulation and enhances the ability of LLMs to understand the consistency and correlation of information within audio sequences. 
We also demonstrate the effectiveness of scaling datasets in an audio separation task: we use large-scale data (36.5k hours), including speech, music, and sound, to train a universal target audio separation model that is not limited to a specific domain. 
Experiments show that UniSep achieves competitive subjective and objective evaluation results compared with single-task models. 
\end{abstract}


\section{Introduction}

Audio source separation separates individual sources within an audio mixture~\cite{7492604,8683007}. 
In recent years, audio source separation has shown impressive performance in various separation tasks, such as speech separation~\cite{wang2018supervised, luo2019conv}, sound separation~\cite{8937253, tzinis2020improving, wisdom2020unsupervised, wang2023consistent}, and music source separation~\cite{defossez2019music, rouard2023hybrid}.
The majority of previous studies focus on source-specific tasks, such as separating distinct speakers in speech separation and isolating different music sources in music separation.
To accomplish these source-specific tasks, models specific to each source must be trained on datasets designed for the respective task~\cite{pons2023gass}.

In contrast to source-specific separation tasks, some research has proposed universal source separation~\cite{8937253, wisdom2021s, chen2022zero}. 
One such approach is target audio source separation, which aims to isolate target audio from mixtures. 
For instance, chen~\textit{et al.}~\cite{chen2022zero} use weakly-labeled data and query-based learning for target audio separation, but their focus remains solely on sound.
Most existing other methods for universal source separation depend on supervised learning techniques and utilize limited training datasets~\cite{8937253, tzinis2020improving, wisdom2021s}. 
Considering the extensive variety of sounds in daily life, these datasets are limited to accurately represent the majority of audio sources~\cite{pons2023gass}. 
Recently, GASS~\cite{pons2023gass} utilizes large-scale supervised datasets to train a model that is capable of separating various types of audio in blind source separation tasks. 
However, a universal solution for target audio separation has not yet been explored. Most approaches typically focus on separating specific domains and sources. They are unable to simultaneously separate a broader range of types, including speech, music, and sound events.

In this study, we propose to build a universal target audio separation model (UniSep) performed on unlimited source domains and unlimited source numbers, which meets the following research problems: (1) how to model the complex audio signal in one unified model, especially, different types of audio commonly have distinct patterns, such as frequency span. (2) whether a unified target audio separation is promising and beneficial compared to previous specific models. (3) whether large-scale audio data can be used to improve the performance of the target audio separation task. We will answer these questions in this study. The main contributions of this study are summarized as follows: 
\begin{itemize}

\item 
Instead of modeling the audio data in spectrogram or waveform, we use a pre-trained audio codec model \cite{soundstream} to convert the audio signal into discrete tokens and model the audio data in a discrete latent space. By using language models with
large-scale parameters \cite{borsos2023audiolm, yang2023uniaudio}, we formulate the traditional audio separation problem as sequence-to-sequence modeling, improving the simplicity of audio separation. 
\item We propose a novel pre-training method that utilizes large-scale audio-only data, reducing costly data simulation requirements. 
This method aims to enhance the ability of the LLMs to understand the consistency and correlation of information within audio sequences, which builds the foundation for the separation tasks.
\item In addition, we scale up the training dataset to 36.5k hours. This enables us to harness the language model's power to handle various source types, ultimately achieving universal target audio separation.
\item 
To the best of our knowledge, we offer the first study on LLM-based universal audio separation. Experimental results show that the UniSep model is not only capable of performing the universal separation task effectively but can also be fine-tuned with a minimal amount of data to accomplish other related tasks, \textit{e.g.} language-queried audio source separation task.

\end{itemize}

\begin{figure*}[t]
    \centering
    \includegraphics[width=\textwidth]{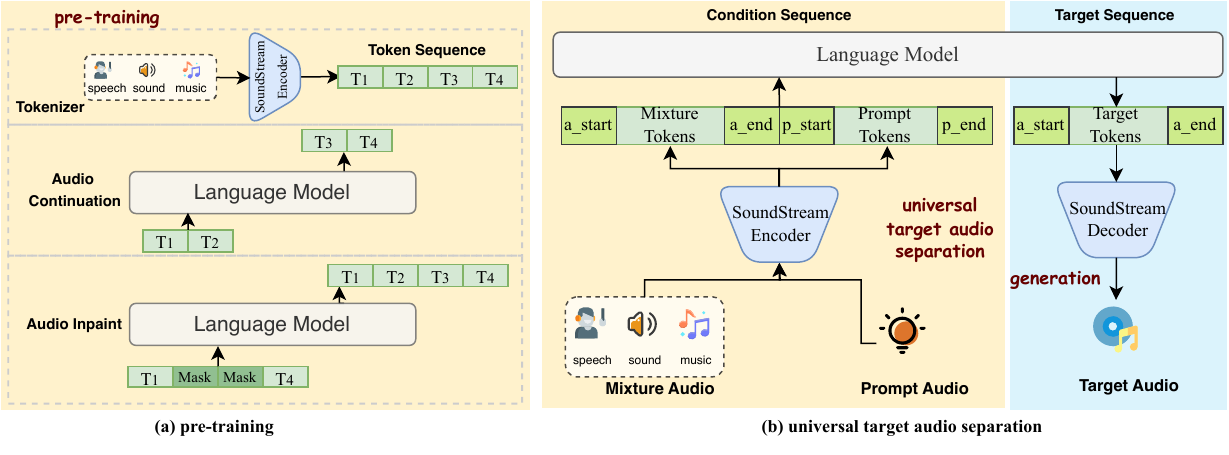}
    \vspace{-25pt}
    \caption{(a) gives the illustration of two pre-training tasks. (b) shows the Sequence layout for UniSep. We encode all input into the discrete token space so that we can directly use language model architectures for audio separation.} 
    \label{fig:uniaudio}
    \vspace{-15pt}
\end{figure*}

\section{Method}
In this section, we present the details of UniSep. We first introduce how to quantize the audio signal into discrete tokens, then we introduce how to use a language model for target audio separation. Lastly, we introduce the proposed pre-training strategy.



\subsection{Audio Tokenization}
In this work, we try to separate any type of audio (speech, sound, music) in a unified discrete latent space. To obtain such latent space, we opt for neural audio codec models \cite{soundstream}. Specifically,  
Assuming an audio signal of duration $d$ with sample rate $f_s$ can be represented by a sequence $\mathbf{x} \in [-1,1]^{d*f_s}$.  An audio neural codec intends to compress ${\mathbf{x}}$ and then recover it as $\mathbf{\hat{x}}$ using an encoder-decoder module with a quantizer:
\begin{equation}
    \mathbf{h} = \text{E}(\mathbf{x}) \in \mathcal{R}^{T*L}; \quad 
    \mathbf{\hat{h}} = \text{Q}(\mathbf{h});
    \quad 
    \mathbf{\hat{x}} = \text{D}(\mathbf{\hat{h}})
\end{equation}
where $T$ denotes the number of audio frames after down-sampling in the encoder, and $L$ denotes the feature dimension of the encoder. $E$, $D$, and $Q$ represent the Encoder, Decoder and Quantizer. Given any frame of encoder output $\mathbf{h}_t$, the token sequence $\mathbf{z}_{t} = [z_t^1, ..., z_t^{N_q}]$ is generated by Residual Vector Quantization (RVQ) \cite{soundstream}, where $N_q$ denotes the number of vector quantization layers. With the discrete representation $\mathbf{z}_t$, $\mathbf{\hat{h}}_t$ is reconstructed as a close estimation of $\mathbf{h}_t$, then $\mathbf{h}_t$ is used to recover $\mathbf{x}_t$ with the decoder. 
For an audio sample with $T$ frame, its discrete representation $\mathbf{z} \in \mathbf{Z}^{T\times N_q}$ is a matrix. In this work, we convert it into a sequence before being processed by the language model: we use a simple flatten operation along the $N_q$ dimension. In this work, we set $N_q=3$.

\subsection{Language Models for Target Audio Separation}
The neural audio codec model allows us to operate on discrete audio representations. We train an autoregressive (AR) decoder-only language model for a target audio separation task. The target audio separation task aims to separate the target audio from a mixture of audio based on the query prompt. In general, previous works \cite{wang2023consistent,chen2022zero} use a query encoder to extract a global embedding to guide the separation. In this study, we directly use the discrete tokens of query prompt to guide the separation. Specifically, we first use the neural audio codec model to extract the discrete tokens for mixture audio, query prompt, and target audio. Then we combine them as one sequence in order, and apply the language model to it, as Figure \ref{fig:uniaudio} (b) shows. 
Then we can train the language model in two ways. The first is building a prefix-language model (prefix-LM), which regards the prompt sequence and mixture sequence as the condition to predict the tokens of target audio, it can be formulated as:
\begin{equation}
\label{AR1}
    p (\mathbf{a} | \mathbf{m}, \mathbf{c}; \theta) = \prod_{t=0}^{T} p(\mathbf{a}_{t}|\mathbf{a}_{<t}, \mathbf{c}, \mathbf{m}; \theta)
\end{equation} 
where $\mathbf{a}$ denotes the target audio tokens, $\mathbf{m}$ and $\mathbf{c}$ denote the mixture sequence and prompt audio tokens. $T$ denotes the sequence length of target tokens. $\theta$ denotes the model parameters. The second way is to build a causal language model (causal-LM) \cite{radford2018improving}, which tries to model the whole sequence in the training process \footnote{Note that causal-LM and prefix-LM have the same model configuration, the difference is attention mask strategy and loss calculation.}. In the inference, we also regard the prompt and mixture as the condition to predict the target sequence:
\begin{equation}
\label{AR2}
    p (U; \theta) = \prod_{t=0}^{N} p(\mathbf{u}_{t}|\mathbf{u}_{<t}; \theta)
\end{equation} 
where $U= [\mathbf{m},\mathbf{c},\mathbf{a}]$, $N$ denotes the total length of the combined sequence. In this work, we adopt the causal-LM for the target audio separation task. We have two considerations: (1) From the training aspect, causal-LM increases the training difficulty and improves the generalization due to it asks LM to model the relationships internally between condition tokens. (2) Our preliminary experimental results show that causal-LM performs better than prefix-LM. 
In the model structure aspect, we adopt a multi-scale transformer \cite{yang2023uniaudio} for the language model. To help the language model better understand the query, mixture, and target sequence, we add the special token [a\_start] and [a\_end] encompass the mixture and target sequence, [p\_start] and [p\_end] encompass the prompt sequence.  

\subsection{Pre-training}

In the previous section, we introduce how to train a causal language model with the triple data (mixture audio, prompt audio, target audio). However, such a training strategy costs a lot of effort to prepare the large-scale dataset. In this part, we introduce a pre-training strategy to utilize a large-scale audio-only dataset, as Figure \ref{fig:uniaudio} (a) shows. Our motivation is that consistency and relevance are the key points for a target audio separation model. \textit{e.g.} The model should be able to output consistent target source audio without any other noise. Furthermore, the model should output the most relevant source audio with the prompt prompt, so the model should learn to understand the most relevant source with the prompt. To improve the consistency and relevance of the language model, we design two pre-training tasks: Audio Continuation and Audio Inpaint.
Specifically, for any audio samples, we extract its token sequence using the audio codec model, and then we design two tasks: \\
\textbf{Audio Continuation} We design an audio continuation pre-training task to improve the ability of the language model to generate consistent audio. We let the LM predict the following audio sequence based on the previous part. \\
\textbf{Audio Inpaint} We design an audio inpaint pre-training task to improve the ability of the language model to understand the relevance between different audio segments. We add a [MASK] token and randomly replace some tokens in the original sequence (with probability 20\% to mask). Then, we use the masked sequences as conditions, to predict the masked token and recover the original sequence.

\section{Experimental Setup}

\subsection{Dataset}
\label{sec:dataset}
As shown in Table~\ref{res_dataset}, our model uses five publicly available datasets with a total duration of 36.5k hours. 
We utilize the LibriLight~\cite{kahn2020libri} and AudioSet~\cite{gemmeke2017audio} for pre-training and employ the MLS~\cite{pratap2020mls}, AudioSet~\cite{gemmeke2017audio}, and MUSDB18~\cite{rafii2017musdb18} for training the audio separation task.

\begin{table}[t]
\centering
\caption{The training dataset used in UniSep. Audio-only denotes we use the original audio without any data simulation. Simulation denotes we come multiple audio as one mixture audio.}
\begin{tabular}{lccc}
\toprule[1pt]
\multirow{2}{*}{\textbf{Dataset}}
& \multirow{2}{*}{\textbf{Source type}}
& \multicolumn{2}{c}{\textbf{Duration(hrs)}}
\\
&
& audio-only & simulation
                 \\   \hline
LibriLight~\cite{kahn2020libri}                 & \multirow{2}{*}{Speech}                        & 20k                  & -                                        \\
MLS~\cite{pratap2020mls}                 &                         & -                  & 10k                                        \\
AudioSet~\cite{gemmeke2017audio}                 & Sound                         & 5.8k                  & 10.1k                                       \\

MUSDB18~\cite{rafii2017musdb18}                & Music                        & -                  & 722                                        \\

\bottomrule[1pt]
\end{tabular}
\label{res_dataset}
\vspace{-10pt}
\end{table}

\noindent \textbf{Pre-training.} 
In our experiments, we utilize the unsupervised LibriLight datasets for pre-training, which are more easily accessible than labeled data. 
For pre-training, we divide the data into ~10-second audio segments, using a total of 20k hours from LibriLight and 5.8k hours from AudioSet. This pre-training takes about 2 weeks. 

\noindent \textbf{Speech.}
For our target speech separation task, we randomly mixed two distinct segments from different speakers, with a -5 to 10 SNR. Each mixed and target speech lasts between 3-10 seconds. We also extract 3 seconds from other target speaker segments as prompt speech.
In this way, we simulate approximately 10k hours of MLS for the target speech separation task. 
We test the models’ performance on the test set of the Libri2Mix test set~\cite{cosentino2020librimix}, sampled at 16 kHz.

\noindent \textbf{Sound.}
AudioSet is a weakly labeled dataset, which has complex caption annotations of each audio clip but lacks accurate timestamps. 
For target audio separation, Some researchers\cite {chen2022zero} pointed out that due to the complexity of audio clips, it is difficult to find a semantically consistent but different audio sample to serve as the query audio. Therefore, we utilize similar methods to simulate data in the AudioSet.
We follow the pipeline proposed in~\cite{chen2022zero}, which uses a Sound Event Detection (SED) system to detect each audio clip and extract more accurate sound segments.
Next, we simulate the dataset using following ways:
(1) We create mixtures by randomly selecting two audio clips with different sound events and simply using the target source in the mixture as prompt audio. (2) We use the SED to detect 4-second audio segments, with the first 2 seconds as prompt and the subsequent 2 seconds as target. 
We simulate approximately 10.1k hours of mixed data, with a signal-to-noise ratio (SNR) ranging from -5 to 10 in both cases.

\noindent \textbf{Music.} MUSDB18 has four different genres along with their isolated drums, bass, vocals, and others stems.
In our approach, we randomly select a 3-second prompt from a specific music track, followed by 5-8 seconds designated as the target. 
Subsequently, we randomly choose 1-2 stems from the other three categories to mix. 
We simulate 722 hours as the train set and 1.44 hours as the test set in target music separation.

\begin{table}[t]
\centering
\caption{Performance evaluation on target speech separation. }
\setlength{\tabcolsep}{1.8mm}
\begin{tabular}{ccccc}
\toprule[1pt]
\multirow{2}{*}{\textbf{Model}}
& \multirow{2}{*}{\textbf{Pre-train}}
& \multicolumn{3}{c}{\textbf{Libri2Mix test set$(\uparrow)$}}
\\
& 
& \multicolumn{1}{c}{\textbf{PESQ}}
& \multicolumn{1}{c}{\textbf{DNSMOS}}
& \multicolumn{1}{c}{\textbf{MUSHRA}}
                    \\ \hline
SpeakerBeam~\cite{vzmolikova2019speakerbeam}           & \multirow{4}{*}{N}        & \textbf{2.89}                  & 3.18                    & 3.68$\pm$0.10             \\
VoiceFilter~\cite{wang2018voicefilter}            &                            & 2.41                  & 3.35                    & 3.43$\pm$0.09            \\
UniAudio(single)~\cite{yang2023uniaudio}       &                            & 1.97                  & 3.93                    & 3.58$\pm$0.08            \\
UniAudio~\cite{yang2023uniaudio}                &                            & 1.88                  & 3.96                    & 3.72$\pm$0.06            \\ \hline
UniSep(Single)               & N                         & 1.85                  & 3.89                    &  3.54$\pm$0.13                    \\
UniSep(Single)                 & Y                        & 2.09                  & 3.96                    &       3.71$\pm$0.14               \\
UniSep                 & N                         & 2.14                  & 4.01                    &          3.89$\pm$0.11            \\
\textbf{UniSep}                 & Y                        & 2.23                  & \textbf{4.07}                    &          \textbf{4.11$\pm$0.11}            \\

\bottomrule[1pt]
\end{tabular}
\label{tab:res_speech}
\vspace{-10pt}
\end{table}

\subsection{System Implementation}
\label{sec:system}
\noindent \textbf{Implementation details.}
Our UniSep is trained with AdamW optimizer ($\beta_1$=0.9, $\beta_2$=0.95, eps=1e-8) with an initial learning rate of 5e-4. 
We perform the same learning rate decay strategy with ~\cite{vaswani2017attention} and the warm-up step is 10000.
Our batch scale is 4800 tokens, which represents the sum of sequence length. The transformer architecture in our UniSep comprises 12 layers for global transformer and 4 layers for local transformer, with each layer containing 8 attention heads. The embedding dimension for the transformer is set to 1536. Our model has a total of 535 million parameters and is trained on 8 Tesla V100 GPUs.
We use a pre-trained neural audio codec model\footnote{https://github.com/yangdongchao/UniAudio} to extract discrete tokens, which is based on SoundStream. 
To eliminate the impact of the codec, we use the audio reconstructed by the codec as the reference audio during evaluation.

\noindent \textbf{System evaluation.}
To measure signal quality, our UniSep is evaluated using both objective and subjective metrics.
Perceptual Evaluation of Speech Quality (PESQ) and non-intrusive Deep Noise Suppression Mean Opinion Score  (DNSMOS)~\cite{reddy2021dnsmos} are signal-level objective quality metrics derived from human auditory research. 
ViSQOL (Virtual Speech Quality Objective Listener)~\cite{hines2015visqol, chinen2020visqol} is an objective metric for assessing perceived audio quality. 
It compares the spectro-temporal similarity between a reference and a test speech signal to generate a MOS-LQO (Mean Opinion Score - Listening Quality Objective) score, which ranges from 1 (worst) to 5 (best). 
STFT distance~\cite{kumar2023highf} measures the loss between the log magnitude spectrograms of the reference and ground truth waveforms. 
We use PESQ and DNSMOS as objective metrics for evaluating speech, and ViSQOL and STFT distance as objective metrics for assessing music and sound. 
Note that since our UniSep uses autoregressive sampling at their output, metrics such as PESQ and scale-invariant signal-to-noise ratio improvement (SI-SNRi) that rely on exact sample alignment between references and estimates are not suitable~\cite{erdogan2023tokensplit}. 
Moreover, we also perform a subjective test to measure the performance of our models, which can provide a more reliable measurement of relative quality.
We conduct a MUltiple Stimuli with Hidden Reference and Anchor (MUSHRA) listening test which is designed to assess the perceived quality of different audio samples under specific conditions.
We use the MUSHRA as the subjective test that primarily evaluates the similarity between generated audio and target audio (ground truth). 
We randomly selected 50 sentences from the test set and recruited 20 people to evaluate the audio. The score range is from 1 to 5. 

\section{Results and analyses}
\begin{table}[htb]
\centering
\renewcommand{\arraystretch}{1.15}
\setlength{\tabcolsep}{1.0mm}
\caption{Performance evaluation on target sound separation.}
\begin{tabular}{lcccc}
\toprule[1pt]
\multirow{2}{*}{\textbf{Model}}
& \multirow{2}{*}{\textbf{Pre-train}}
& \multicolumn{2}{c}{\textbf{AudioSet test set}}
\\
& 
& \multicolumn{1}{c}{\textbf{STFT$(\downarrow)$}}
& \multicolumn{1}{c}{\textbf{ViSQOL$(\uparrow)$}}
& \multicolumn{1}{c}{\textbf{MUSHRA$(\uparrow)$}}
                    \\ \hline
    CLAPSep~\cite{ma2024clapsep} & N & 4.13 & 2.99 & -
    \\
    SoloAudio~\cite{helin2024soloaudio} & N & 2.86 & 3.33 & -
    \\
    USS-ResUNet30~\cite{kong2023universal} & N & 2.28 & 3.31 & 3.2$\pm$0.27
    \\
UniSep(Single)                 & N                         & 1.43                  &3.86 & 2.95$\pm$0.15                                        \\
UniSep(Single)                 & Y                        & 1.35           &3.94       & 3.24$\pm$0.13                                        \\
UniSep                 & N                         & 1.28            &4.02      & 3.55$\pm$0.10                                       \\
\textbf{UniSep}                 & Y                        & \textbf{1.24}         &\textbf{4.10}         & \textbf{3.73$\pm$0.08}                                        \\

\bottomrule[1pt]
\end{tabular}
\label{tab:res_sound}
\end{table}

\begin{table*}[t]
\centering
\renewcommand{\arraystretch}{1.1}
\caption{Performance evaluation on target music separation(MUSDB18 test set).  avg represents the average evaluation results across all categories. STFT means STFT distance.}
\begin{tabular}{lcccccccccccc}
\toprule[1pt]
\multirow{2}{*}{\textbf{Model}}
& \multirow{2}{*}{\textbf{Pre-train}}
& \multirow{2}{*}{\textbf{STFT$(\downarrow)$}}
& \multicolumn{5}{c}{\textbf{ViSQOL$(\uparrow)$}}
& \multicolumn{5}{c}{\textbf{MUSHRA$(\uparrow)$}}
\\ \cmidrule(r){4-8}  \cmidrule(r){9-13}
& & &
vocals & drums & bass & other & avg &  vocals & drums & bass & other & avg

                    \\ \hline
    CLAPSep~\cite{ma2024clapsep}                 & N                        & 5.78& - & -&-&-                & 3.60     
&   -    
&-
&-
& -
&- \\
    SoloAudio~\cite{helin2024soloaudio}                 & N                         & 2.90& - & -&-&-                & 3.58     
&   -    
&-
&-
& -
&- \\
UniSep(Single)                 & N                         & 1.82& 3.99&4.03&4.08&4.04                & 4.04      
&   3.16$\pm$0.10     
&2.84$\pm$0.14
&2.86$\pm$0.14
& 3.03$\pm$0.13
&2.97 \\
UniSep(Single)                & Y                        &  1.63 & 4.17&4.06&4.12&4.08               & 4.11   
&   3.52$\pm$0.14
& 3.03$\pm$0.12
& 3.14$\pm$0.18
& 3.26$\pm$0.11  
& 3.24                \\
UniSep                 & N                         & 1.33 & 4.22&4.19&4.18&4.21                & 4.20   
&       3.74$\pm$0.09
&3.41$\pm$0.09
&3.33$\pm$0.09
&3.60$\pm$0.07
&3.52
\\
\textbf{UniSep}                 & Y                        & \textbf{1.21} & \textbf{4.26}&\textbf{4.22}&\textbf{4.19}&\textbf{4.24}                 & \textbf{4.23}                    &     \textbf{3.98$\pm$0.11}
& \textbf{3.55$\pm$0.07}
&\textbf{3.50$\pm$0.08}
& \textbf{3.85$\pm$0.12}
&\textbf{3.72} \\

\bottomrule[1pt]
\end{tabular}
\label{tab:res_music}
\vspace{-17pt}
\end{table*}

\begin{figure*}[tb]
    \centering
    \includegraphics[width=\textwidth]{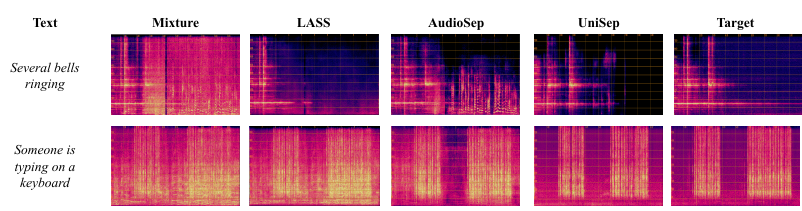}
    \vspace{-20pt}
    \caption{Visualization results on language-queried audio source separation.} 
    \label{fig:language_res}
    \vspace{-15pt}
\end{figure*}

UniSep is first trained on the segmented audio-only LibriLight dataset with about 20k hours.
Next-token prediction and mask techniques accomplish audio continuation and audio inpainting, which also enhance the language model's comprehension capabilities. 
Based on the pre-trained model, we train the target audio separation task on the simulated MLS, AudioSet, ESC-50, and MUSDB18 datasets. 
Ultimately, we can use a universal model to perform universal audio separation for speech, sound, and music. UniSep (Single) represents the training of each task using our model structure in Table~\ref{tab:res_speech}~\ref{tab:res_sound}~\ref{tab:res_music}, so there are three separate single-task models. We can see that single-task models are poor than unified model, which validated the importance of building a unified separation model.
In the following, we evaluate our UniSep on speech, sound, and music test set. 

\subsection{Results on Speech}
\label{sec:res_on_speech}
For speech separation, we use the Libri2Mix test set to evaluate the target speech separation task.
As shown in Table~\ref{tab:res_speech},  our UniSep performs better than the single-task model in both objective metrics and subjective evaluations. We can also observe that the model performance is improved with pre-training, which validates the effectiveness of the pre-training methods we proposed. Furthermore, compared with previous works, such as UniAudio, VoiceFilter, and SpeakerBeam, UniSep also shows better performance, \textit{e.g.} better DNSMOS and MUSHRA metrics. We also observe a similar finding with previous work \cite{erdogan2023tokensplit}: the objective metric PESQ is may not suitable for a sampling-based generation model, due to it depending on the strict alignment between reference and generated speech. 

\subsection{Results on Sound}
In this section, we evaluate our UniSep model on the AudioSet test set.
From the Table~\ref{tab:res_sound}, we can conclude the same results with Section~\ref{sec:res_on_speech}. 
UniSep with pre-training significantly outperforms different baselines across multiple metrics, validating UniSep’s superiority. 
Due to time and resource constraints, we performed only demo-level comparisons (average) on CLAPSep~\cite{ma2024clapsep} and SoloAudio~\cite{helin2024soloaudio}. 
Their weaker performance could be due to a lack of specialized training on music datasets. 

Additionally, it's important to mention that we used the AudioSet dataset in this experiment, which includes speech, music, and sound events, with speech and music making up the majority. Therefore, during the training and inference stages, any combination of these three types of data could exist. In the AudioSet section of our demo page, one can listen to examples of music mixed with speech or sound events, which also demonstrates the universality of our UniSep.

\subsection{Results on Music}
As demonstrated in Table~\ref{tab:res_music}, we explore the separation performance of different stems in MUSDB18.
Overall, the subjective and objective metrics of the three experiments demonstrate that our method can generate audio with better perceptual quality. The average metrics indicate that the results are essentially consistent with both speech and sound. 
This suggests that universal multitasking outperforms single-task approaches, and the pre-training method we employed effectively enhances the model's performance.


\subsection{UniSep as foundation model for downstream task}
UniSep can be fine-tuned using a small amount of data to accomplish other tasks, \textit{e.g.} language-queried audio source separation task \cite{liu2022separate, liu2023separate}. We show that UniSep can be fine-tuned on text-audio pairs, and then it can be used to separate audio based on text description. 
We utilize T5~\cite{2020t5} to encode the text description.
The encoded text is then used as a query to guide the separation process, allowing UniSep to isolate the audio sources that match the given textual description. 

Table~\ref{tab:res_language} shows the results, where our UniSep, after fine-tuning, performs slightly worse compared to AudioSep.
It is worth noting that AudioSep is trained on a massive dataset of 14,000 hours of audio-text pairs. 
In contrast, our UniSep is fine-tuned on a much smaller dataset of only around 16.7 hours from the AudioCaps dataset. 
Compared to AudioSep, UniSep achieves 96.94\% of ViSQOL performance with only 0.12\% audio-text pairs. 
Under identical data conditions, compared to LASS, UniSep achieved a relative performance improvement of 36.09\% in STFT, demonstrating its efficient performance for downstream tasks.
We also visualize the spectrograms for audio mixtures, target audio sources, and separated sources using different models, as depicted in Figure~\ref{fig:language_res}. 
The spectrogram patterns of the separated sources resemble those of the target sources, which highlights the model’s effectiveness. Audio samples can be found on our project page.

\begin{table}[t]
\centering
\renewcommand{\arraystretch}{1.15}
\caption{Performance evaluation on language-queried audio source separation. 
`Hours' refers to the volume of the audio-text pairs dataset used during the training process. 
UniSep has the same amount of audio-text data as LASS.}
\begin{tabular}{lccc}
\toprule[1pt]
\multirow{2}{*}{\textbf{Model}} &
\multirow{2}{*}{\textbf{Hours}}
& \multicolumn{2}{c}{\textbf{AudioCaps test set}}
\\
&
& \multicolumn{1}{c}{\textbf{STFT$(\downarrow)$}}
& \multicolumn{1}{c}{\textbf{ViSQOL$(\uparrow)$}}
                    \\ \hline
    AudioSep~\cite{liu2023separate}    
    & 14k & \textbf{1.97}                  & \textbf{3.60}                                                 \\ \hline
                    LASS~\cite{liu2022separate}                                   & 17.3  & 3.63                  & 3.40    \\

UniSep+Fine-tuning                                  &16.7   & 2.32         &3.49                                              \\

\bottomrule[1pt]
\end{tabular}
\label{tab:res_language}
\vspace{-15pt}
\end{table}


\section{Conclusion}
We propose a universal target audio separation model, UniSep, based on language models. Different from previous separation models that try to separate the audio in continuous space \textit{e.g.} spectrogram, we propose to separate the audio in a discrete latent space. Then we propose to use language model for target audio separation task. We also propose a novel pre-training strategy to use large-scale audio-only datasets, which significantly reduces the efforts of data simulation. 
To the best of our knowledge, UniSep is the first unified model that can separate any audio source without considering the audio types and the number of audio sources. 
Extensive experiments show the effectiveness of UniSep, we demonstrate the effectiveness of scaling data in audio separation. 
In the future, we can improve UniSeq from two aspects: (1) Training a better audio tokenizer; (2) Scaling the model size and data quantity. 

\vspace{50pt}

\bibliographystyle{IEEEbib}
\bibliography{icme2025references}


\end{document}